\documentclass[10pt]{article}
\usepackage[letterpaper]{geometry}
\usepackage{hicss}
\usepackage{times}
\usepackage[none]{hyphenat}
\usepackage{url}
\usepackage{latexsym}

\usepackage{algorithm}
\usepackage{algorithmic}
\usepackage{amsmath}
\usepackage{amsfonts}
\usepackage{bm}
\usepackage{tabularx}
\usepackage[subtle]{savetrees}

\usepackage{amssymb}

\newcommand{\qed}{\ensuremath{\blacksquare}}

\usepackage{graphicx}
\graphicspath{{images/}}
\title{Occam's Razor in Residential PV-Battery Systems: Theoretical Interpretation, Practical Implications, and Possible Improvements}

\author{Mostafa Farrokhabadi \\
University of Calgary \\
{\underline{mostafa.farrokhabadi@ucalgary.ca}}}

\date{June 10, 2024}

\begin{document}
\maketitle
\begin{abstract}
This paper presents a theoretical interpretation and explores possible improvements of a widely adopted rule-based control for residential solar photovoltaics (PV) paired with battery storage systems (BSS). The method is referred to as Occam's control in this paper, given its simplicity and as a tribute to the 14\textsuperscript{th}-century William of Ockham. Using the self-consumption-maximization application, it is proven that Occam's control is a special case of a larger category of optimization methods called online convex learning. Thus, for the first time, a theoretical upper bound is derived for this control method. Furthermore, based on the theoretical insight, an alternative algorithm is devised on the same complexity level that outperforms Occam's. Practical data is used to evaluate the performance of these learning methods as compared to the classical rolling-horizon linear/quadratic programming. Findings support online learning methods for residential applications given their low complexity and small computation, communication, and data footprint. Consequences include improved economics for residential PV-BSS systems and mitigation of distribution systems' operational challenges associated with high PV penetration.

\end{abstract}

\subsubsection*{Keywords:}

Online Convex Learning, Residential Energy Management Systems, Energy Arbitrage, Peak Shaving

\section{Introduction}

The adoption of solar photovoltaic (PV) systems in residential settings has been on an upward trend for the past decade \citep{SOVACOOL2022112868}. The trend is primarily driven by technological advancements, declining costs, and environmental-friendly mandates and awareness. For example, \citep{AZUATALAM2019555} reports that the residential PV costs have dropped from roughly
12.5 \$/W in 2006 to approx. 2.42 \$/W in 2016. It is added the Australian Energy Market Operator (AEMO) projects an annual cost decline of 1.5\% until 2040. Saying that, doubts are being cast on whether the momentum in penetration of residential PV can be maintained \citep{KHEZRI2022111763}. First, the decline in Feed-in-Tarrifs (FiTs) in areas with already high penetration of PV systems has negatively impacted the residential PV investment rate of return. For example, countries such as Germany and the UK have reduced their FiTs \citep{DONG2020112330}, and Switzerland has eliminated it \citep{LI2023106763}. Second, high penetration of residential PVs creates well-studied operational challenges due to the inherent uncertainty and negative/low net demand leading to overvoltages, e.g., see \citep{6194986}. Finally, attempts have been made to legislate regulatory frameworks impeding solar PV projects in some jurisdictions. Though not passed ultimately, the Florida Senate Bill 1024 is an example of such a regulatory framework, proposing to reduce the benefits of net metering \citep{fl_senate_bill_1024}.

It is intuitive that pairing residential PV with battery storage systems (BSS), such as lithium-ion batteries, enhances the reliability and efficiency of residential energy supply. In fact, the increasing electricity price in jurisdictions such as Ontario and Alberta in Canada combined with decreasing FiTs enhances the financial gain of PV-BSS duos for residential consumers \citep{DOLTER2018135}. The combination allows homeowners to store excess solar energy generated during the day for use during nighttime or periods of low sunlight, thereby reducing dependence on the grid and lowering electricity bills \citep{HOPPMANN20141101}. Furthermore, the integration of PV and storage systems contributes to a more balanced net demand profile, mitigating operational issues associated with a high penetration of PV. Thus, despite the decrease in FiTs, favorable subsidies and regulatory frameworks are expected to further incentivize behind-the-meter BSS \citep{UDDIN201712}. Hence, the residential layer of the electrical grid is becoming increasingly penetrated by PV-BSS systems \citep{CHREIM2024250}.

Energy management systems (EMS) for residential PV-BSS are evolving rapidly; yet for many jurisdictions and within the foreseeable future, such strategies will likely remain focused on local applications that do not require coordination among these systems. This limitation is due to the extensive need for communication infrastructure, the lack of proper regulation and incentive mechanisms, and the requirement for Advanced Metering Infrastructure (AMI) to facilitate coordination between EMS and centralized operation centers \citep{AZUATALAM2019555, BEAUDIN2015318}. Consequently, the two primary applications for these systems are self-consumption maximization (SCM) and time-of-use arbitrage (TA) \citep{azuatalam2018techno}. Interestingly, both strategies benefit homeowners and help utilities by decreasing peak demand, thus enhancing grid operation reliability. 

A wide range of EMS methods are proposed for both SCM and TA applications for residential PV-BSS, e.g., see \citep{MOSHOVEL2015567, abdulla2016optimal, 6338330, CASTILLOCAGIGAL20112338, NYHOLM2016148, LUTHANDER201580}. These can be broadly categorized into two main methods, namely rule-based, referred to as Occam's control in this paper \footnote{In philosophy, Occam's Razor advises against more complex solutions in the presence of simpler approaches with a smaller set of components \citep{walsh1979occam}. This is known in science as the ``principle of parsimony" \citep{Schaffer}. Thus, given its simplicity, Occam's control is named after William of Ockham, who argued in favor of the principle.}, and optimization-based \citep{AZUATALAM2019555}. Occam's controls consist of rule-based if-then logics and are easy to design and implement; they require the lowest computational, communication, and data storage resources. On the other hand, depending on the objective \footnote{It is discussed later in Section 4 that Occam's control is the global optimal policy for a certain objective in the presence of perfect forecast.}, they ``theoretically" result in suboptimal performance. Optimization-based methods include mixed-integer programming, artificial intelligence, dynamic programming, and heuristics. These methods share different degrees of complexity and exhibit various levels of optimality gap. Shared among optimization-based methods is the need for time series forecasting and thus extensive historical operational data, proprietary solvers or inference engines, and a sizeable need for data storage capacity \citep{BEAUDIN2015318}. Such requirements are typically challenging to satisfy in the residential systems, suppressing these methods' practical implementation. Hence, many commercial products for residential EMS rely on a variation of Occam's controls \citep{AZUATALAM2019555}. Furthermore, thanks to low computational complexity, Occam's controls are the standard operational strategy for long-term planning studies, given that more complex techniques may not be computationally feasible for planning time horizons \citep{9937170}. 

Despite the widespread use, to the authors' knowledge, no theoretical interpretation of Occam's control has been provided to date.  Furthermore, there have been no theory-inspired efforts to devise superior methods on the same complexity level, at least not in the context of residential PV-BSS. Motivated by these gaps, this paper proves that Occam's control is a subset of online convex learning. Building on this insight, possible improvements are explored at the same computational and data storage complexity level. An alternative data-driven learning algorithm is proposed that is highly sample-efficient, i.e., rapidly converging to its long-term average performance. The proposed method does not rely on time series prediction and instead dispatches BSS based on the most recent observations. Furthermore, the formulation is ideal for ensemble or hybrid implementations. Thus, the control method reported here can be widely adopted by providing a feasible and superior alternative to Occam's control with no additional complexity. The improved performance reduces the investment rate of return of residential PV-BSS by increasing the efficiency of planning and operation, leading to an increased share of clean energy in the residential sector. In addition, more effective BSS operation mitigates the distribution systems' challenges associated with the increased share of residential PV. We name the proposed method momentum-optimized smart (MOS) control. Thus, the contributions of this paper are as follows:

\begin{itemize}
    \item A theoretical interpretation of Occam's control in residential PV-BSS applications, providing a performance upper bound for the first time.
    \item MOS, a momentum-optimized online convex learning for EMS of residential PV-BSS that outperforms Occam's controls on the same computational, communication, and data resources complexity.
    \item An analysis of these methods' performance compared to rolling-horizon quadratic programming, leveraging real-world measurements.
\end{itemize}

The rest of this paper is organized as follows: Section 2 provides the problem statement; it also formulates Occam's and quadratic programming frameworks. Section 3 describes the proposed online learning method. Section 4 presents simulation results and analysis. Section 5 concludes the paper and lays out future works. 

\section{Problem Statement and Benchmarks}
\begin{figure}[t!]
    \centering
	\includegraphics[width=0.75\linewidth]{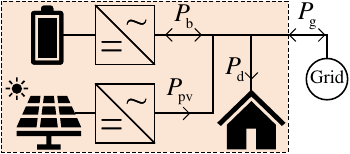}
	\caption{Residential PV-BSS structure.}
	\label{fig: 1}       % Give a unique label
\end{figure}

The structure of the residential PV-BSS system is shown in Figure \ref{fig: 1}. The shaded area demonstrates the electrical boundaries of the residential system. In this figure, $P_\mathrm{b}$ is the ac power injected/absorbed by the BSS, $P_\mathrm{pv}$ is the ac power injected by PV, $P_\mathrm{d}$ is the ac power consumed by the residential unit, and $P_\mathrm{g}$ is the net power consumed/produced by the residential system, i.e.:
\begin{equation}
P_{\mathrm{g},t}=P_{\mathrm{d},t}+P_{\mathrm{b},t}-P_{\mathrm{pv},t}
\label{eq:0}
\end{equation}
where $P_{\mathrm{b},t}$ is positive for battery charging and is negative for battery discharging. Similarly, $P_{\mathrm{g},t}$ is positive for net power consumption by the residential system, and is negative for net power production.

Without the loss of generality, the problem is formulated for the SCM application. The TA application shares a similar form of objectives and constraints, hence, it is not discussed here. Two control methodologies, Occam's control and quadratic programming, are formulated.

\subsection{Optimization Formulation}
For $x \in \mathbb{R}^n$, define the $\mathcal{L}_2(x)$ norm as $\|x\|_2 = \sqrt{\sum_{i=1}^n x_t^2}$, where $x_t$ is the t\textsuperscript{th} entry of the vector $x$ for $t \in \mathcal{T}$. The problem can be formulated in the optimization framework as follows:

\begin{equation}
\min_{P_\mathrm{g}}: \quad \|P_{\mathrm{g}}\|_2^2, \quad \forall t \in \mathcal{T} \\
\label{eq:1}
\end{equation}
such that: 
\begin{equation}
\mathrm{P_b^{min}} \le P_{\mathrm{b}, t} \leq \mathrm{P_b^{max}}
\label{eq:2}
\end{equation}
\begin{equation}    
\begin{aligned}
& e_t=e_{t-1}+\eta P_{\mathrm{b}, t}\Delta t, \ \ \ \ \text{if}\quad P_{\mathrm{b}, t}\ge 0\\
& e_t=e_{t-1}+P_{\mathrm{b}, t}\Delta t /\eta, \ \ \text{if}\quad P_{\mathrm{b}, t}<0\\
\end{aligned}
\label{eq:3}
\end{equation}
\begin{equation}
\mathrm{e^{min}} \le e_t \leq \mathrm{e^{max}}
\label{eq:4}
\end{equation}
where $\mathcal{T}$ is the set of time intervals, $\mathrm{P_b^{min}}$ and $\mathrm{P_b^{max}}$ are the minimum and maximum level of BSS power, $e_t$ is the state of energy at the end of time interval $t$, $\eta$ is the charging/discharging efficiency, and $\mathrm{e^{min}}$ and $\mathrm{e^{max}}$ are the minimum and maximum state of charge (SoC), respectively. 

Equation \ref{eq:1} is the problem objective function, i.e., minimizing the residential power exchange with the grid. The choice of the square of $\mathcal{L}_2$ norm is motivated by its differentiability, as compared to, e.g., $\mathcal{L}_1=\|x\|_1 = \sum_{i=1}^n |x_i|$. Furthermore, this objective function formulation fits the framework for benchmark quadratic programming solvers and eliminates the need for piece-wise linear formulation. Finally, $\mathcal{L}_2^2$ implicitly penalizes the net demand peak, as we show later in Section 4. Equation \ref{eq:2} ensures the decision variable $P_{\mathrm{b}, t}$ stays within acceptable ranges. Equation \ref{eq:3} describes the evolution of the BSS state of energy through time. Equation \ref{eq:4} ensures that the BSS state of energy remains within acceptable ranges.

\subsection{Occam's Control}

In this method, the dispatch $P_\mathrm{b}$ is calculated based on the most recent observation of the PV power $P_\mathrm{pv}$ and the demand $P_\mathrm{d}$. Hence, if there is a surplus PV power after meeting the demand, the battery is set to charge, otherwise, it discharges. The charge and discharge decisions are subject to the availability of enough energy capacity. Algorithm 1 describes Occam's control.

\begin{algorithm}[t!]
\caption{Occam's Control}
\label{alg:self-consumption}
\begin{algorithmic}[1]
\STATE \textbf{Input:} $\mathrm{P_b^{max}}$, $\mathrm{P_b^{min}}$, $\mathrm{e^{max}}$, $\mathrm{e^{min}}$, $\eta$
\STATE Initialize set of time intervals $\mathcal{T}$
\STATE Initialize $P_{\mathrm{b}, 0}=0$, $e_0=0.5\ \mathrm{e^{max}}$
\FOR{$t \in \mathcal{T}$}
    \STATE \textbf{Input:} $P_{\mathrm{d}, t-1}$, $P_{\mathrm{pv}, t-1}$
    \STATE $P_{\mathrm{r}, t} = P_{\mathrm{pv}, t-1}-P_{\mathrm{d}, t-1}$
    \IF{$P_{\mathrm{r}, t} \ge 0$} 
        \STATE $e_{t}=e_{t-1}+\eta P_{\mathrm{r}, t} \Delta t$
        \IF{$e_{t} \le \mathrm{e^{max}}$}
            \STATE $P_{\mathrm{b}, t}=P_{\mathrm{r}, t}$
        \ELSE
            \STATE $P_{\mathrm{b}, t}=(\mathrm{e^{max}}-e_{t-1})/(\eta\Delta t)$
        \ENDIF
        \STATE $e_{t} = e_{t-1} + \eta P_{\mathrm{b}, t}\Delta t$
    \ELSE
        \STATE $e_{t}=e_{t-1}+(P_{\mathrm{r}, t} \Delta t)/\eta$
        \IF{$e_{t} \ge \mathrm{e^{min}}$}
            \STATE $P_{\mathrm{b}, t}=P_{\mathrm{r}, t}$
        \ELSE
            \STATE $P_{\mathrm{b}, t}=(\mathrm{e^{min}}-e_{t-1})\eta/\Delta t$
        \ENDIF
        \STATE $e_{t} = e_{t-1} + P_{\mathrm{b}, t}\Delta t/\eta$
    \ENDIF
    \STATE $P_{\mathrm{g}, t}=P_{\mathrm{d}, t}+P_{\mathrm{b}, t}-P_{\mathrm{pv}, t}$
\ENDFOR
\STATE Report (\ref{eq:1})
\end{algorithmic}
\end{algorithm}

\subsection{Rolling Horizon Mixed Integer Quadratic Programming (MIQP)}
The problem described in Section 2.1 can be solved by rolling horizon MIQP solvers, such as CPLEX and Gurobi, see \citep{CPLEX, Gurobi}. In this case, the optimization problem is explicitly solved at each interval over a horizon, for example, the next 24 hours. Thus, the method relies on time series prediction. Algorithm 2 describes this control methodology. In this algorithm, $\hat{P}_{\mathrm{x}, t}$ is the forecasted power during interval $t$. Note that for the sake of notational consistency, there are no explicit integer variables in algorithm 2; however, the ``if" statements are logically coded by the integer decision variables on if the battery is charging ($P_{\mathrm{b}, t} \ge 0$) or discharging ($P_{\mathrm{b}, t} < 0$). 

\begin{algorithm}[t!]
\caption{MIQP Control}
\label{alg:self-consumption}
\begin{algorithmic}[1]
\STATE \textbf{Input:} $\mathrm{P_b^{max}}$, $\mathrm{P_b^{min}}$, $\mathrm{e^{max}}$, $\mathrm{e^{min}}$, $\eta$
\STATE Initialize set of time intervals $\mathcal{T}$
\STATE Initialize $P_{\mathrm{b}, 0}=0$, $e_0=0.5\ \mathrm{e^{max}}$
\FOR{$t \in \mathcal{T}$}
    \STATE \textbf{Input:} $\hat{P}_{\mathrm{d}, t}$, $\hat{P}_{\mathrm{pv}, t}$, \ldots, $\hat{P}_{\mathrm{d}, t+24}$, $\hat{P}_{\mathrm{pv}, t+24}$
    \STATE Solve (1-4) for $t, \ldots, t+24$
    \IF{$P_{\mathrm{b}, t} \ge 0$} 
    %    \STATE $e_{t}=e_{t-1}+\eta P_{\mathrm{b}, t} \Delta t$
    %    \IF{$e_{t} > \mathrm{e^{max}}$}
    %        \STATE $P_{\mathrm{b}, t}=(\mathrm{e^{max}}-e_{t-1})/(\eta\Delta t)$
    %    \ENDIF
        \STATE $e_{t} = e_{t-1} + \eta P_{\mathrm{b}, t}\Delta t$
    \ELSE
    %    \STATE $e_{t}=e_{t-1}+(P_{\mathrm{b}, t} \Delta t)/\eta$
    %    \IF{$e_{t} < \mathrm{e^{min}}$}
    %        \STATE $P_{\mathrm{b}, t}=(\mathrm{e^{min}}-e_{t-1})\eta/\Delta t$
    %    \ENDIF
        \STATE $e_{t} = e_{t-1} + P_{\mathrm{b}, t}\Delta t/\eta$
    \ENDIF
    \STATE $P_{\mathrm{g}, t}=P_{\mathrm{d}, t}+P_{\mathrm{b}, t}-P_{\mathrm{pv}, t}$
\ENDFOR
\STATE Report (\ref{eq:1})
\end{algorithmic}
\end{algorithm}
   
\section{Occam's Control Theory and the Proposed MOS}
\subsection{Online Convex Learning}

\textbf{Definition 1} The set $C \subseteq \mathbb{R}^n$ is convex if and only if  $\forall x, y \in C$ and $\forall \lambda \in [0, 1], \lambda x + (1-\lambda) y \in C.$

\textbf{Definition 2} A function \( f: \mathbb{R}^n \rightarrow \mathbb{R} \) is said to be convex over a convex set \( C \subseteq \mathbb{R}^n \) if, for all \( x, y \in C \) and for all \( \lambda \in [0, 1] \), the following inequality holds:
\[ 
f(\lambda x + (1-\lambda) y) \leq \lambda f(x) + (1-\lambda) f(y)
\]

Based on Definitions 1-2, an online convex learning problem is defined as follows \citep{zinkevich2003}:
\begin{enumerate}
    \item A convex feasible set \( C \subseteq \mathbb{R}^n \).
    \item An infinite sequence of convex cost functions \( \{ f_t: C \rightarrow \mathbb{R} \}_{t=1}^{\infty} \).
\end{enumerate}
At each time step \( t \):
\begin{enumerate}
    \item The algorithm selects a point \( x_t \in C \) without knowledge of the cost function \( f_t \).
    \item The cost function \( f_t \) is then revealed, and the algorithm incurs a cost \( f_t(x_t) \).
\end{enumerate}

The objective of online convex learning is to minimize the cumulative cost over time, i.e.,
\begin{equation}
\min_{x \in C}\quad \sum_{t=1}^T f_t(x_t)
\label{eq:5}
\end{equation}

\textbf{Lemma 1} The problem described by Equation \ref{eq:1}-\ref{eq:4} satisfies the online convex learning.

\textbf{Proof} Set $f_t(x_t)=(P_{\mathrm{g}, t})^2$ and hence $x_t=P_{\mathrm{b}, t}$. Convexity of $f_t$ is given. At each time step $t$, $P_{\mathrm{b}, t}$ is selected from the following convex feasibility set\footnote{The feasibility set in \citep{zinkevich2003} remains unchanged in time. This doesn't impact our proofs, since our problem's feasibility set, though changing in time, satisfies all the framework's assumptions. It, however, impacts the performance evaluation, as discussed later in this paper.}:
\begin{equation}
\begin{aligned}
S_t=[&\min(\mathrm{P_b^{min}}, (\mathrm{e^{min}}-e_{t-1})\eta/\Delta t),\\
&\max(\mathrm{P_b^{max}}, (\mathrm{e^{max}}-e_{t-1})/(\eta\Delta t))]
\end{aligned}
\label{eq:6}
\end{equation}
The function $f_t$ is then revealed by Equation \ref{eq:0}.\qed

\subsection{Theoretical Framework}

The algorithm proposed in this paper is based on the \textit{Greedy Projection} algorithm proposed by in \citep{zinkevich2003}. They discuss seven assumptions as prerequisites of their algorithm, as follows.

Define the distance \( d(x, y) = \|x - y\| \). Thus, in the context of our problem, it is assumed that:
\begin{enumerate}
    \item \( \exists\ N \in \mathbb{R} \) such that \( d(x, y) \leq N \), \(\ \forall x, y \in S_t \),
    \item Consider the sequence \( \{x^1, x^2, \ldots\} \in S_t \); if there exists a point \( x \in \mathbb{R}^n \) such that \( x = \lim_{k \to \infty} x^k \), then \( x \in S_t \).
    \item \(\exists\ x \in S_t \).
    \item The cost function \( f_t \) is differentiable \(\forall t \in \mathcal{T} \).
    \item \(\exists\ N \in \mathbb{R} \) such that \( \|\nabla f_t\| \leq N \), \(\forall t \in \mathcal{T} \) and \(\forall x \in S_t \). 
    \item \( \nabla f_t \) is computable given \( x \in S_t \) and \(\forall t \in \mathcal{T} \).
    \item \( \arg\min_{x \in S_t} d(x, y) \) is computable \(\forall y \in \mathbb{R}^n \). Thus, define \( \mathcal{P}_{S_t}(y) = \arg\min_{x \in S_t} d(x, y) \).
\end{enumerate}

\textbf{Lemma 2} The problem described by Equations \ref{eq:1}-\ref{eq:4} satisfies the above assumptions.

\textbf{Proof} Assumptions 1-3 implies the that $S_t$ is bounded, closed, and non-empty. These are given by Equation \ref{eq:6}. Assumptions 4 and 6 are satisfied given that $\nabla f_t = 2P_{\mathrm{g}, t},\ \forall t \in \mathcal{T}$. Assumption 5 is satisfied so long that $P_{\mathrm{d}, t}-P_{\mathrm{pv}, t}$ is bounded, which is given. Assumption 7 is proven as follows; let $S^{\mathrm{min}}_t=\min(\mathrm{P_b^{min}}, (\mathrm{e^{min}}-e_{t-1})\eta/\Delta t)$ and $S^{\mathrm{max}}_t=\max(\mathrm{P_b^{max}}, (\mathrm{e^{max}}-e_{t-1})/(\eta\Delta t))$. Thus:
\begin{equation}
\mathcal{P}_{S_t}(y_t) =
\begin{cases} 
      S^{\mathrm{min}}_t & \text{if } y_t \leq S^{\mathrm{min}}_t \\
      y_t & \text{if } S^{\mathrm{min}}_t < y_t \leq S^{\mathrm{max}}_t \\
      S^{\mathrm{max}}_t & \text{if } y_t > S^{\mathrm{min}}_t
   \end{cases}
\label{eq:7}
\end{equation}\qed

The Greedy Projection online convex learning method proposed in \citep{zinkevich2003}, applied to the problem described in Equations \ref{eq:1}-\ref{eq:4}, is as follows:

\begin{enumerate}
    \item Select an arbitrary $P_{\mathrm{b}, 0} \in S_0$;
    \item Select an arbitrary sequence of $\alpha_1, \alpha_2, ...\in \mathbb{R}^+$;
    \item Calculate $P_{\mathrm{b}, t}=\mathcal{P}_{S_t}(P_{\mathrm{b}, t-1}-\alpha_{t-1} \nabla f_{t}(P_{\mathrm{b}, t-1}))$.
\end{enumerate}

\textbf{Lemma 3} Algorithm 1 is a special case of the Greedy Projection algorithm.

\textbf{Proof} Let $\alpha_t=0.5, \forall t \in \mathcal{T}$. Given that $\nabla f_{t}(P_{\mathrm{b}, t})=2P_{\mathrm{g}, t}$ and that $P_{\mathrm{g}, t}=P_{\mathrm{d}, t}+P_{\mathrm{b}, t}-P_{\mathrm{pv}, t}$, the calculation in step 3 of Greedy Projection is the same as step 6 in Algorithm 1 projected on the feasibility set described in steps 7-23 in Algorithm 1.\qed

A notable consequence of Lemma 3 is that it allows for a theoretical analysis of Occam's Control performance. 

\textbf{Definition 3} An algorithm's ``Regret" is defined by:
\[
R_T = \sum_{t=1}^{T} f_t(x_t) - \min_{\mathbf{x} \in C} \sum_{t=1}^{T} f_t(\mathbf{x})
\]
where the second term is the optimal decision in hindsight, given it remains fixed in time.

It is proven in \citep{zinkevich2003} that the Greedy Projection algorithm's regret with $\alpha_t=t^{-1/2}$ tends to zero, i.e., $\textrm{lim sup}_{T \to \infty} R_T/T \leq 0$. This is with the assumption that the feasibility set $C$ is fixed in time. Relaxing the constraint in Equation \ref{eq:4}, the feasibility set described in Equation \ref{eq:6} is modified to $P=[\mathrm{P_b^{min},P_b^{max}}]$. Thus, it can be implied that the upper bound of performance of the Greedy Project in our problem is $\sum_{t=1}^{T} f_t(x^*)$, where $\mathbf{x}^* = \arg\min_{\mathbf{x} \in P} \sum_{t=1}^{T} f_t(\mathbf{x})$. Note that this is an optimistic upper bound, given that we relaxed the problem.

\subsection{Proposed Algorithm}

Based on the insight obtained from Lemma 3, a method is proposed in this Subsection by introducing a BSS action momentum term to $f_t(x_t)$, as follows:
\begin{equation}
    f'_t(x_t)=(P_{\mathrm{g}, t})^2+\mu e^{(-|P_{\mathrm{b}, t}|)}
    \label{eq:obj}
\end{equation}
where $\mu$ is a tunable hyperparameter. At each time step $t$, $P_{\mathrm{b}, t}$ is selected from the following convex feasibility set:
\begin{equation}
S'_t =
\begin{cases} 
      [S^{\mathrm{min}}_t,0] & \text{if } P_{\mathrm{b}, t-1} < 0\\
       S_t & \text{if } P_{\mathrm{b}, t-1} = 0 \\
      [0,S^{\mathrm{max}}_t] & \text{if } P_{\mathrm{b}, t-1} > 0
   \end{cases}
\label{eq:set}
\end{equation}

\textbf{Lemma 4} $f'_t(x_t)$ is convex over $S'_t$.

\textbf{Proof} $\nabla e^{(-|P_{\mathrm{b}, t}|)}$ is $-\mathrm{sgn}(P_{\mathrm{b}, t})e^{(-|P_{\mathrm{b}, t}|)}$, where $-\mathrm{sgn}(P_{\mathrm{b}, t})$ is the sign function. Thus, the gradient is expressed as follows:
\begin{equation}
\nabla  e^{(-|P_{\mathrm{b}, t}|)} = 
\begin{cases} 
-e^{(-P_{\mathrm{b}, t})} & \text{if } P_{\mathrm{b}, t} > 0, \\
e^{(P_{\mathrm{b}, t})} & \text{if } P_{\mathrm{b}, t} < 0, \\
0 & \text{if } P_{\mathrm{b}, t} = 0 \text{ (subgradient)}
\end{cases}
\label{eq:grad}
\end{equation}
Convexity of Equation \ref{eq:grad} is given in each of the three $P_{\mathrm{b}, t}$ ranges. Thus, given that the $S'_t$ is defined such that $P_{b,t}$ cannot change sign compared to the decision in the previous interval, $f'_t(x_t)$ does not encounter discontinuities or non-linear transitions that would affect its convexity.\qed

\begin{figure}[t!]
    \centering
	\includegraphics[width=0.85\linewidth]{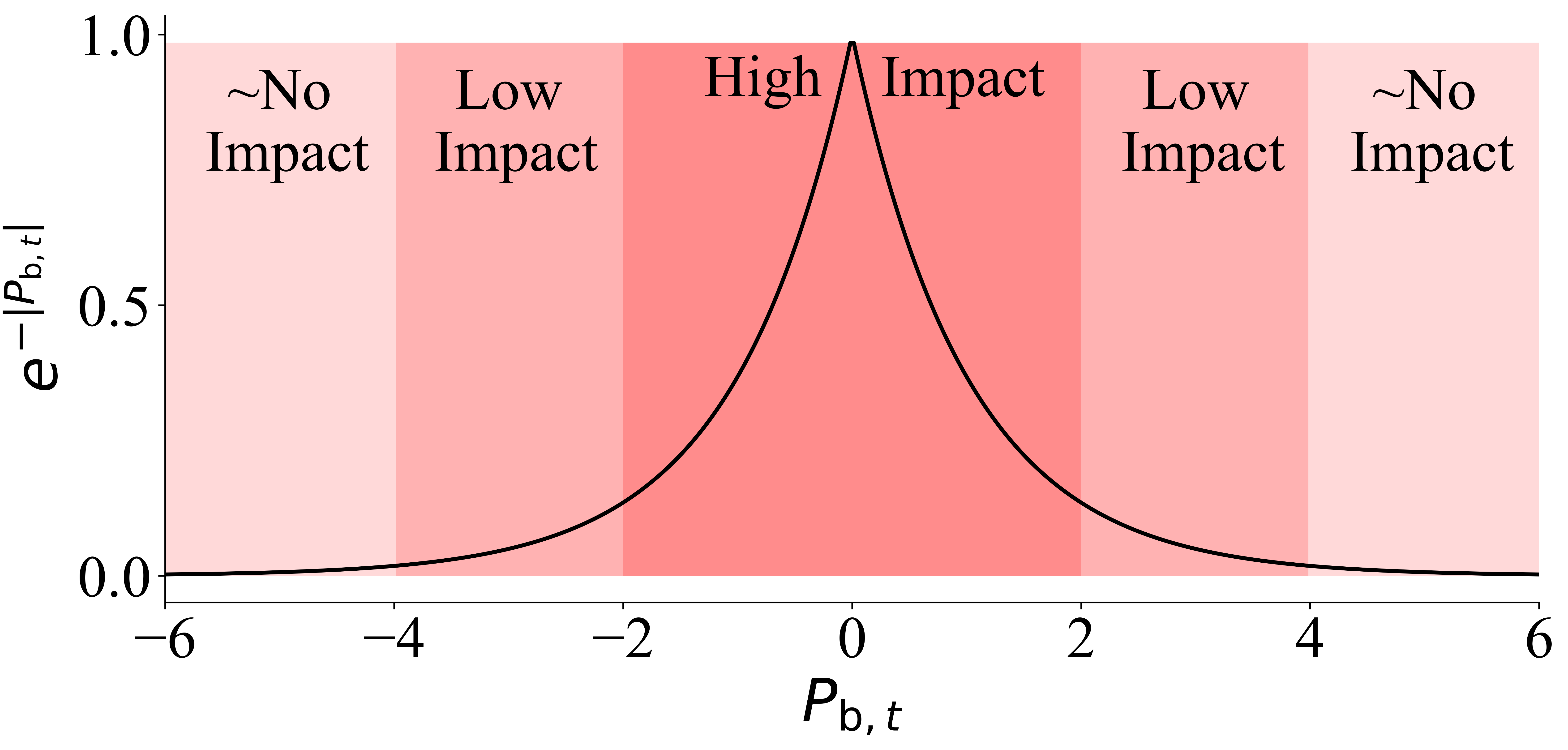}
	\caption{$e^{(-|P_{\mathrm{b}, t}|)}$ function.}
	\label{fig:exponent}       % Give a unique label
\end{figure}

The proof for the rest of the assumptions discussed in Section 3.2 remains unchanged. Thus, MOS is described in Algorithm 3. In this algorithm, the decision at each interval $P{\mathrm{b}, t}$ is updated based on the gradient of the modified objective function in Equation \ref{eq:obj}. The intuition behind adding the exponent can be explained by referring to Figure \ref{fig:exponent}. Thus, the exponent serves as a moment of inertia for the BSS actions, i.e., when $P_{\mathrm{b}, t-1}$ is slightly positive (charging) or negative (discharging), the next action tends to be more charge or discharge, respectively. The moment of inertia exponentially decreases as the battery charging or discharging setpoint increases. The intuition is that the upward/downward trends of the net demand profile are captured more effectively. Furthermore, as discussed before, the feasibility set in Equation \ref{eq:set} does not allow a sudden change of charge to discharge and vice versa between two consecutive intervals. $\alpha$ is the learning rate, i.e., the magnitude of the steps taken toward the gradient descent. Introducing $\kappa$ is another modification compared to the Greedy Projection algorithm. $\kappa$ is a regularization factor between the decision variable and that of 24 steps before, i.e., the similar time interval of the day before. The regularization is inspired by the well-studied daily seasonality in the residential demand and PV \citep{9316723}.

\begin{algorithm}[t!]
\caption{MOS Algorithm}
\label{alg:self-consumption}
\begin{algorithmic}[1]
\STATE \textbf{Input:} $\mathrm{P_b^{max}}$, $\mathrm{P_b^{min}}$, $\mathrm{e^{max}}$, $\mathrm{e^{min}}$, $\eta$ 
\STATE \textbf{Input:} $\alpha$, $\mu$, $\kappa$ \COMMENT{hyperparameters}
\STATE Initialize set of time intervals $\mathcal{T}$
\STATE Initialize $P_{\mathrm{b}, 0}=0$, $e_0=0.5\ \mathrm{e^{max}}$
\FOR{$t \in \mathcal{T}$}
    \STATE \textbf{Input:} $P_{\mathrm{d}, t-1}$, $P_{\mathrm{pv}, t-1}$, $P_{\mathrm{b}, t-(24/\Delta t)+\Delta t}$
    \STATE $\nabla f'_{1, t-1} = 2P_{\mathrm{g}, t-1}$
    \STATE $\nabla f'_{2, t-1} = \nabla  e^{(-|P_{\mathrm{b}, t}|)}$ \COMMENT{defined in Equation \ref{eq:obj}}
    \STATE $\nabla f'_{t-1}=\nabla f'_{1, t-1}+\mu \nabla f'_{2, t-1}$
    \STATE $P^*_{\mathrm{b}, t}=(1-\kappa)P_{\mathrm{b}, t-1}-\alpha\nabla f'_{t-1}+\kappa P_{\mathrm{b}, t-(24/\Delta t)+\Delta t}$
    \STATE $P_{\mathrm{b}, t}=\mathcal{P}_{S'_t}(P^*_{\mathrm{b}, t})$
    \IF{$P_{\mathrm{b}, t} \ge 0$} 
        \STATE $e_{t} = e_{t-1} + \eta P_{\mathrm{b}, t}\Delta t$
    \ELSE
        \STATE $e_{t} = e_{t-1} + P_{\mathrm{b}, t}\Delta t/\eta$
    \ENDIF
    \STATE $P_{\mathrm{g}, t}=P_{\mathrm{d}, t}+P_{\mathrm{b}, t}-P_{\mathrm{pv}, t}$
\ENDFOR
\STATE Report (\ref{eq:1})
\end{algorithmic}
\end{algorithm}

\section{Results}

\subsection{Data}
\begin{figure*}[t!]
    \centering
	\includegraphics[width=1\textwidth]{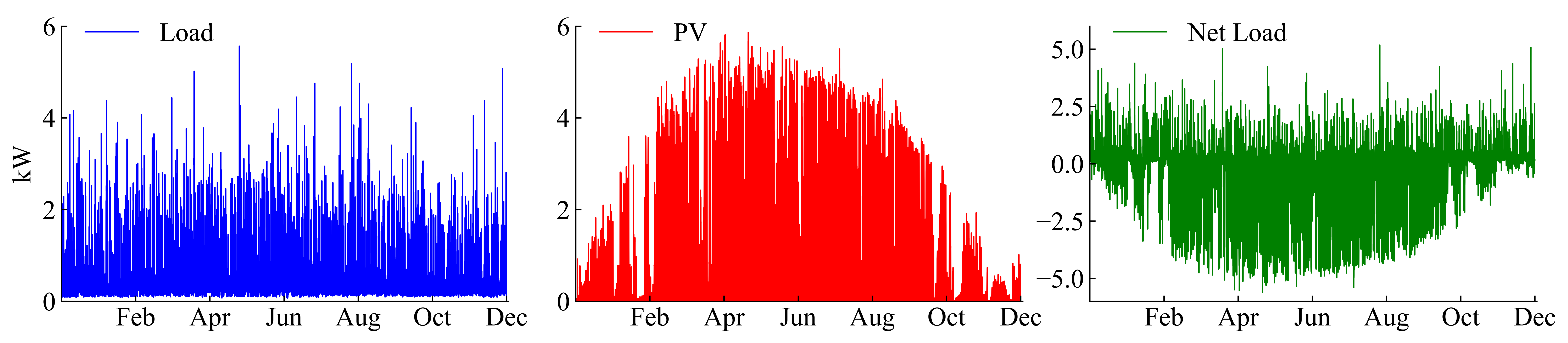}
	\caption{Annual 30-minute profiles for residential load and PV.}
	\label{fig:2}       % Give a unique label
\end{figure*}

One year of real-world residential load and PV data is used to compare Algorithms 1, 2, and 3. The load data is obtained from \citep{MVLVNetworks}, which includes anonymized residential demand profiles in 30-minute granularity, operating under AusNet Services' jurisdiction in Australia. The PV profiles are obtained in 5-minute granularity from solar panels implemented on a university campus in North America. These profiles, as well as the ``net demand," are shown in Figure \ref{fig:2}. The authors acknowledge that selecting the profiles from two different locations may compromise the fidelity of the net demand profile; saying that there is no evidence that such a compromise has a clear impact on the fidelity of the conclusions.

As seen in Figure \ref{fig:2}, net demand barely becomes negative in the first and last two months. Hence, the algorithms' performance for these months is irrelevant since the BSS barely charges. Thus, to have a more effective comparison, the algorithms are tested on the profile from March to October. Moreover, the BSS power and energy capacity are varied through trainings and tests to analyze sensitivity to BSS ratings. Without the loss of generality, the charging/discharging efficiency is considered 1. If anything, this assumption helps reduce the complexity of the conventional optimization problem, eliminating the need for integer variables. The assumption has no tangible impact on the computational complexity of Occam's method or MOS. 

\subsection{Conventional Optimization Setup}

As seen in Algorithm 2, the rolling horizon method optimizes the BSS operation over a 24-hour-ahead horizon, using the persistence method for forecasting the load and PV profiles for the next 24 hours \citep{ZHANG2018343}. Persistence is selected for two main reasons; first, it is simple to implement and does not rely on extensive historical profiles, making it attractive for practical residential applications. Second, the persistence method has been shown to be a relatively accurate predictor for short-term forecasting \citep{9739082}. The optimization problem is coded in Pyomo \citep{Bynum2021}, and solved using CPLEX \citep{CPLEX}.

\subsection{MOS Training}

MOS needs hyperparameter tuning for $\alpha$, $\mu$, and $\kappa$. Thus, a data-driven grid-search approach is proposed here that tunes the hyper-parameters that minimize:
\begin{equation}
\|P_{\mathrm{g}}\|_2^2+\gamma \|P_{\mathrm{g}}\|_1, \quad \forall t \in \mathcal{T}
\label{eq:8}
\end{equation}
where $\| P_{\mathrm{g}} \|_1 = \sum_t |P_{\mathrm{g}, t}|$ ($\mathcal{L}_1$ norm). Given the tuning's data-driven nature, there is the flexibility to add any other desired penalty (or reward) to Equation \ref{eq:8}. In this case, a small percentage of the $\mathcal{L}_1$ regularization will ensure MOS does not overfit on $\mathcal{L}_2^2$ minimization, thus, $\gamma=0.02$.

MOS shows a considerable sample efficiency property. The intuition is that the net demand profile is stationary over the period; while the magnitude of net demand may change from one period to another, the relation autocorrelation stays the same. The theoretical analysis of such sample efficiency is deferred to future works. In this context, more analysis is needed to understand the performance drift and out-of-distribution generalization capabilities. In this Subsection, the algorithm is trained only on the month of March and tested on April-October. The tuning is done using the hyperparameter optimization framework Optuna \citep{Akiba2019}, with $\alpha \in [0.01, 1.0]$, $\mu \in [0, 5]$, and $\kappa \in [0,0.75]$. The ranges are selected experimentally by monitoring the values to which the tuning method converges. The tuning result for various BSS sizes is shown in Table \ref{tab:1}.

\begin{table*}[t!]
    \centering
    \begin{tabular}{c| c c c c c c c c c}
     \textbf{kW-kWh} & \textbf{2-4} & \textbf{2-8} & \textbf{2-12} & \textbf{4-8} & \textbf{4-16} & \textbf{4-24} & \textbf{6-12} & \textbf{6-24} & \textbf{6-36} \\
     \hline\hline
    $\alpha$ & 0.051 & 0.132 & 0.22 & 0.083 & 0.15 & 0.218 & 0.132 & 0.217 & 0.227 \\
    $\mu$ & 0.236 & 0.761 & 1.743 & 0.876 & 1.87 & 2.3 & 1.775 & 2.293 & 2.3233 \\
    $\kappa$ & 0.128 & 0.278 & 0.526 & 0.262 & 0.463 & 0.518 & 0.442 & 0.517 & 0.5 \\
    \end{tabular}
    \caption{Hyperparameters of MOS.}
    \label{tab:1}
\end{table*}

\begin{figure*}[t!]
    \centering
	\includegraphics[width=1\textwidth]{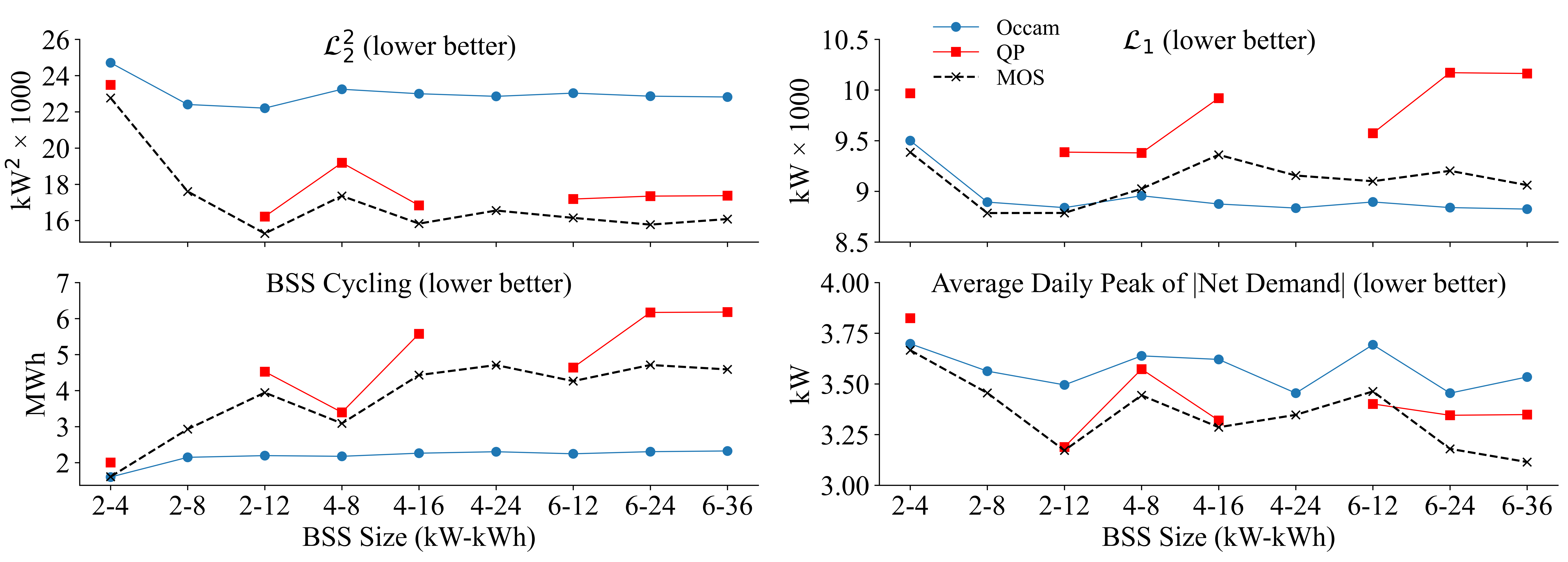}
	\caption{Performance comparisons.}
	\label{fig:3}       % Give a unique label
\end{figure*}

\subsection{Performance Comparison}

 Data shown in Figure \ref{fig:3} is used to evaluate the performance of the control methods. Thus, a test set from April to October is used over nine different BSS sizes. The results for the training set are not reported but are similar in proportion to the results of the test set, except that MOS does even better. The methods' performances are shown in \ref{fig:3}; Occam's control is shown by the blue line with circle marks, the quadratic programming (QP) method by the red line with square marks, and MOS by the dashed black line with cross marks. Four different performance metrics are compared, $\mathcal{L}_2^2$, $\mathcal{L}_1$, BSS cycling, and average daily peak of net demand absolute value. These are compared over the BSS size of 2, 4, and 6 kW, each with the C-rate of 1/2, 1/4, and 1/6. The methods are executed on a Windows PC with 13\textsuperscript{th} Gen Intel(R) Core(TM) i7-13700H 2.90 GHz and 64 GB of RAM. The main observations are as follows:

\begin{itemize}
    \item The data from BSS sizes 2-8 and 4-24 is missing for the LP method. This is because the method did not converge for at least one of the 24-hour-ahead runs\footnote{It is not clear why the method did not converge and the authors did not put a significant effort into investigating the root cause, given it is irrelevant to the paper's narrative. The codes and data will be publicly available on a Git repository for further investigations.}. This demonstrates the complexity of these solvers and the likelihood of computational issues, even in the presence of powerful computation resources.
    \item MOS outperforms LP in all metrics.
    \item MOS outperforms Occam's control in $\mathcal{L}_2^2$, in expense of higher cycling. Occam's slightly outperform MOS in $\mathcal{L}_1$ (variations of $\pm$2.5\%). Furthermore, MOS is more effective in peak reduction.
    \item Occam's outperforms LP in all metrics except $\mathcal{L}_1$.
    \item The benefit of larger BSS diminishes quickly.
\end{itemize}

\subsection{Computation Comparions}

LP takes approximately 200 s to solve the problem for one BSS size over the test period; there is no meaningful difference between Occam's control and MOS, both taking around 0.1 s over the test period. In fact, thanks to low complexity and their rule-based nature, the later controls can be implemented on rudimentary computation resources, making them ideal for residential applications. Regarding data storage capability, the entire data for one year of operation is 137 kB. Thus, all control methods require only a few kB of storage. Note that MOS requires one month of data for training offline. During operation, LP relies on the persistence forecast and thus requires the past 24 hours of data. Similarly, MOS tracks the past 24 hours of data, though it only uses the data for the interval -24h+$\Delta t$.

\subsection{Analysis}

\begin{figure}[t!]
    \centering
	\includegraphics[width=0.85\linewidth]{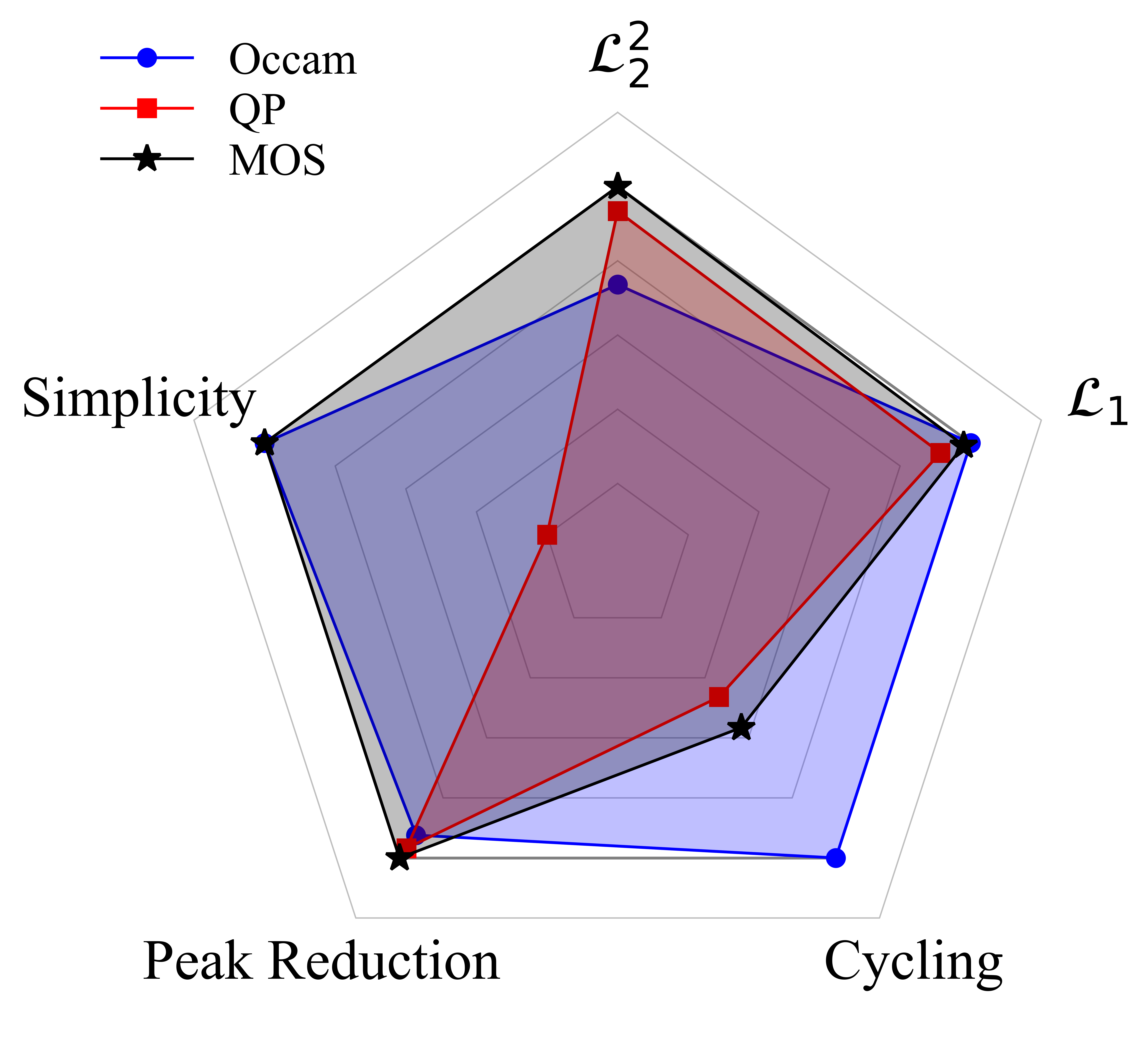}
	\caption{Normalized avg.\ performance (wider better).}
	\label{fig:5}       % Give a unique label
\end{figure}

\begin{figure*}[t!]
    \centering
	\includegraphics[width=1\linewidth]{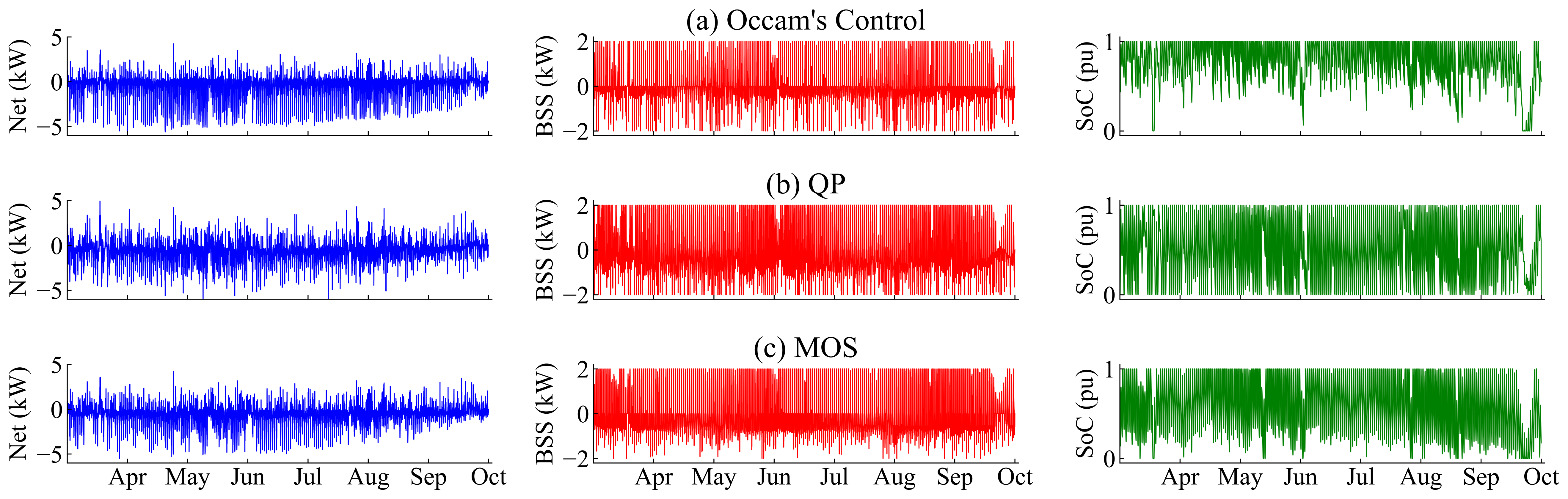}
	\caption{Net demand (blue), BSS power (red), and SoC (green) for BSS size of 2-12.}
	\label{fig:6}       % Give a unique label
\end{figure*}

The normalized average performances over BSS sizes are shown in Figure \ref{fig:5}. Note that the BSS sizes 2-8 and 4-24 are excluded due to LP not converging. All metrics in this figure are quantitative, except simplicity, which is a qualitative metric of complexity and computational/storage intensity. To investigate the differences in algorithms' behavior, the BSS size 2-12 is selected as a representative example. Thus, the algorithms' net demand, BSS power, and SoC profiles over the test set are shown in Figure \ref{fig:6}. 

It can be observed that Occam's control does not effectively use the BSS SoC range, hovering above 0.5 and reaching 1 quite often. This is because the net demand profile is unbalanced toward negative values, as seen in Figure \ref{fig:3}. On the contrary, MOS tends to use a wider range of the BSS capacity, thanks to the additional momentum component, resulting in the battery being discharged more under MOS. The benefit is clear in the $\mathcal{L}_2^2$ norm minimization; while both methods may have similar performance for positive values of net demand, MOS does considerably better on the negative ranges, yielding a considerably lower $\mathcal{L}_2^2$. Thus, it can be argued that the extra cycling under MOS is a favorable tradeoff in situations where the electricity price is asymmetrical between import and export, as well as when the price is nonlinear with respect to the `power demand' (kW). Furthermore, even in the absence of such scenarios, more effective $\mathcal{L}_2^2$ minimization yields a more balanced net demand profile between positive and negative values, mitigating operational issues associated with high penetration of PV, such as overvoltages. In general, more rigorous technical and economic studies on the cycling v.s $\mathcal{L}_2^2$ are needed and are deferred to future works. In addition, in scenarios where the net demand distribution is closer to a Gaussian with a mean of 0, the gap between MOS and Occam's control is expected to close. In other words, MOS' performance is expected to be more robust than Occam's control to net demand distribution parameters. Thus, performance sensitivity to load various load profile distributions can also be a topic for future works. 

The performance difference in $\mathcal{L}_1$ is much smaller, with Occam's control and MOS exhibiting a similar performance, while LP slightly underperforms. Considering the cycling, if $\mathcal{L}_1$ is the metric of choice, Occam's control is superior. This happens in situations where the electricity bill is a linear function of energy (kWh), with symmetrical tariffs for import and export. This is unlikely in practice, as many jurisdictions have a demand component, explicitly penalize higher peaks, or have asymmetrical tariffs. 

Occam's' effectiveness in $\mathcal{L}_1$ minimization is expected. As proven in Section 3.2., Occam's control is a special case of Greedy Projection. This special case, it turns out, is the global optimal policy for $\mathcal{L}_1$ minimization in the presence of a perfect forecast for the next interval. Since $\mathcal{L}_1$ linearly penalize kWh deviation from 0, conserving/releasing BSS capacity for future reward at the expense of immediate penalty does not make sense; there is no kW value in the energy capacity, i.e., the BSS must be ``greedy."

Compared to MOS, LP slightly underperforms in all metrics while being considerably more complex to implement and execute. Saying that, the performance of LP depends on the forecast accuracy and horizon. As discussed before, more advanced forecasting methods require substantial historical data and are complex to productive, making them less suitable for local residential applications.

\section{Conclusions}

This paper evaluated this principle of parsimony for the control of residential PV-BSS systems. For the first time, a theoretical interpretation was provided for a widely adopted simple rule-based method, referred to in this paper as Occam's control, leading to an upper-bound performance analysis. Built on this insight, a data-driven sample-efficient method was proposed that can meaningfully outperform Occam's control without a practical compromise in simplicity. The superiority in $\mathcal{L}_2^2$ minimization has a meaningful impact on the economics of residential PV-BSS systems; using MOS in planning studies can result in more utilized resources, and using it in operation can reduce the supplied cost of electricity and mitigate operational issues associated with high penetration of PV. Simulation studies based on real-world data demonstrated the superiority of simplicity and supported the paper's proposed method. Future works include theoretical analysis of MOS' sample-efficiency, economic analysis of performance metrics, and robustness to net demand profile distribution. Conclusions cannot be applied to other applications and domains without rigorous studies. 

\textbf{Git Repo:} Data and codes will be publicly available upon formal publication.

\bibliographystyle{apalike}
\bibliography{hicss}

\end{document}